\documentstyle[aps]{revtex}

\draft

\tightenlines

\begin{document}

\title{{\small Published in Chaos, Solitons and Fractals, 12(2001)1431-1437} \\
Incomplete statistics and nonextensive generalizations of
statistical mechanics}

\author{Qiuping A. Wang}
\address{Institut Sup\'erieur des Mat\'eriaux du Mans,\\ 44, Avenue F.A.
Bartholdi, 72000 Le Mans, France}

\date{\today}

\maketitle

\begin{abstract}
Statistical mechanics is generalized on the basis of an
information theory for inexact or incomplete probability
distribution. A parameterized normalization is proposed and leads
to a nonextensive entropy. The resulting incomplete statistical
mechanics is proved to have the same theoretical characteristics
as Tsallis one which is based on the conventional normalization.

\end{abstract}



\section{Introduction}
We need a generalized statistical mechanics to replace
Boltzmann-Gibbs-Shannon (BGS) one because BGS theory is not
capable of interpreting some observed results of physical systems
with complicated strong interaction, long range correlation,
longtime memory or with noneuclidean and nonsmooth space-time
structure such as in the theory of fractal space-time and
Cantorian $E^\infty$\cite{Krog00,Nasc00,Nott66,Ahme00}. Examples
of such systems exist everywhere, from cosmic systems, ordinary
optical, magnetic or electronic materials around us, fractal and
chaotic systems, till microscopic systems such as nuclei. For
detailed comments on the breakdowns of BGS theory and on the
possible solutions, the reader is advised to read references
\cite{Tsal95,Tsal99} and the references there-in. Complicated
physical systems are often nonextensive. For this kind of
systems, we already have Tsallis nonextensive statistical
mechanics \cite{Tsal88,Tsal98} which has been successfully used
to interpret some peculiar physical phenomena and fractal
structures. Tsallis entropy is given as follows :
\begin{equation}
S_q= -k\frac{1-\sum_{i=1}^{v}p_i^q}{1-q} , (q \in R)    \label{1}
\end{equation}
where $p_i$ is the probability of the state $i$ among $v$ possible
ones
\begin{equation}
\sum_{i=1}^{v}p_i=1.                    \label{2}
\end{equation}
The independence of two probability distributions
$[p_1(A)...p_i(A)...p_v(A)]$ et $[p_1(B)...p_j(B)...p_u(B)]$ is
defined as usual by\footnote{This kind of ``independence" for
correlated subsystems can be interpreted by the existence of
thermodynamic equilibrium in the composite system\cite{Abe01}}
\begin{equation}
p_i(A)p_j(B)=p_{ij}(A\cup B)                \label{3}
\end{equation}
where $p_{ij}(A\cup B)$ is the probability for the state $i$ and
$j$ to be occupied at the same time. We obviously have
$\sum_{i=1}^v\sum_{j=1}^u p_{ij}(A\cup B)=1$). We see that Tsallis
statistics, as BGS one, is constructed within Kolmogorov algebra
of complete probability distribution \cite{Reny66}. On the other
hand, the expectation value $<O>$ of an observable $\hat{O}$ is
supposed to be calculated in the following way,

\begin{equation}
<O>=\frac{\sum_{i=1}^{v}p_i^qO_i}{\sum_{i=1}^{v}p_i^q}. \label{4}
\end{equation}
where $O_i$ is the value of $\hat{O}$ at the state $i$.

I want to emphasize here that, as mentioned above, up to now, all
statistical theories (BGS, Tsallis, etc) are based on the
information theory for complete probability distribution according
to 'complete random variables' ($CRV$)\cite{Reny66}. A $CRV$,
$\xi$, takes distinct values $X=\{x_1,x_2,...,x_v\}$ (also
referred to as $v$ events or states) with probabilities
$P=\{p_1,p_2,...,p_v\}$ and $X$ constitutes a complete ensemble
$\Omega_x$ defined by all possible values of $\xi$. A simple
example of $CRV$ is the position of a particle in a closed box.
All the possible values of the position constitute a complete
ensemble of positions defined in the box. We call $\xi$ an
independent $CRV$ if all its values are independent and
incompatible (exclusive). In this case, $P$ is called a complete
distribution and satisfies the requirement of Eq. (\ref{2}).

We can say that all statistical theories constructed within
Kolmogorov algebra of complete probability distribution should be
logically applied to physical systems of which all the possible
physical states are well-known and for which we can in practice
count all of the states to carry out the calculation of
probability or of whatever physical quantities. This often
requires that we can find the exact hamiltonian and also the exact
solutions of the equation of motion of the systems to know all
possible states and to obtain the exact values of physical
quantities.

The generalization of BGS theory in the present paper is based on
the information theory for incomplete probability distribution
with incomplete random variables ($IRV$), that is the number $w$
of the possible values $\{x_1,x_2,...,x_w\}$ of $IRV$ is greater
or smaller than $v$. Another case where the present generalization
can be applied is that, though $w=v$, one can not calculate the
exact probability distribution $P=\{p_1,p_2,...,p_v\}$.

In physics, the above two cases are possible if we do not know or
can not write analytically all the interactions in a system. In
these cases, the solution of the equation of motion is no longer
exact, and we can no more in practice count all the possible
states and calculate in a precise way any physical quantity of the
system even for the countable (well-known) states. So the
calculation of the probability $p_i$ is not exact either. Within
this hypothesis, Eq. (\ref{1}) should be written as \cite{Reny66}
\begin{equation}
\sum_{i=1}^{w}p_i=Q\neq 1.                    \label{6}
\end{equation}
where $w$ is only the number of the countable states given by the
solution of the equation of motion and can be greater or smaller
than $v$, the real number of states of the physical system under
consideration. $Q$ should depend on the way to find the countable
states and probabilities, or in other words, on the neglected
interactions.

The idea of the present generalization of BGS statistics is that
we recognize the inadequacy of our knowledge of physical systems,
i.e. Eq.(\ref{6}), and try to tackle the problems in a approximate
way allowing us to work within Kolmogorov algebra of probability
as BGS statistical mechanics does. We also require that the BGS
statistics be a special case of the generalized formalisms. As
mentioned above, the present $nonextensive$ generalization is
carried out with the hypothesis of an incomplete ensemble of
states and of an inexact probability distribution. The reader will
find that this incomplete statistics has the same theoretical
features as Tsallis one for complete ensemble of states.

\section{Incomplete normalization}

Now let us postulate
\begin{equation}
\sum_{i=1}^{w}\frac{p_i}{Q}=\sum_{i=1}^{w}p_i^q \label{7}
\end{equation}
so that
\begin{equation}
\sum_{i=1}^{w}p_i^q=1, (q\in[0,\infty])             \label{8}
\end{equation}
since $p_i<1$, we have to set $q\in[0,\infty]$. Eq. (\ref{4}) now
becomes :
\begin{equation}
<O>=\sum_{i=1}^{w}p_i^qO_i.                          \label{8a}
\end{equation}

Don't forget that the $w$ states are only the well-known
(countable) ones and therefore do not constitute a complete
ensemble of states of the system under consideration. $w$ can be
greater or smaller than the real number of all the possible
states, depending on the approximations we use to find the
analytic expression of hamiltonian and the solution of the
equation of motion of the system.

Eq.(\ref{7}) is a kind of redistribution of the effect of
neglected interactions (or of the unknown events) on the known
events. This is quite normal because the known events and their
probability distribution are closely related to the neglected
interactions. This {\it $q$-deformation} of probability $p_i^q$ is
not a new invention. It was the choice of almost all the authors
who intended to generalize Shannon information theory
\cite{Tsal95,Reny66}. It is used by R\'enyi to calculate the {\it
$q$-order measure of information} and related to an {\it average
gain} of lost information due to the incomplete distribution.
$p_i^q$ is also used in the fractal theory to favor contributions
from events with relatively high values (when $q>1$) or low values
(when $q<1$) in the calculation of multifractal measure
\cite{Fractal}. In any case, Eq. (\ref{8}) is a (reasonable)
simplification of Eq.(\ref{6}) which allows, we will find later in
this paper, to work within the mathematical framework of BGS
statistical mechanics.

In the {\it simplified incomplete normalization} Eq. (\ref{8}),
$p_i^q$ can be called the {\it effective probability} of the event
$i$ and $p_i$ the {\it real probability} which is physically
useful only when $q=1$ for the cases where no interaction is
neglected and $w$ is the total number of events in a complete
ensemble. When $q$ is different from unity, the difference $(q-1)$
is related to the neglected interactions. We can see later in this
paper that $q$ is related to the extra information (entropy)
compared to that we obtain in BGS framework. $q$ can also be
related in explicit ways to other basic quantities such as
internal energy or free energy of the system.

\section{Incomplete information theory}

It is well known that the conventional (extensive) information
theory is based on the following postulates concerning the missing
information $I(N)$ to determine the state of a system $\Omega$ of
N elements\cite{Reny66} :

$n^o$1) $I(1)=0$ (no missing information if there is only one
event)

$n^o$2) $I(e)=1$ (information unity)

$n^o$3) $I(N)<I(N+1)$ (more information with more elements)

$n^o$4) $I(\prod_{i=1}^v N_i)=\sum_{i=1}^v I(N_i)$ (additivity)

$n^o$5) $I(N)=I_v+\sum_{i=1}^v p_iI(Ni)$ (additivity of
information measure in two steps)

where $v$ is the number of all the subsystems $\Omega_i$ with
$N_i$ elements and $p_i=\frac{N_i}{N}$ is the probability to find
an element in $\Omega_i$ (equiprobability). $I_v$ is the missing
information to determine in what subsystem an element will be
found. The postulates $n^o$1 to $n^o$4 lead to Hartley formula of
information measure \cite{Reny66} :
\begin{equation}
I(N)=lnN .                                    \label{9}
\end{equation}
Then the postulate $n^o$5 can yield Shannon formula for $I_v$ and
entropy.

For nonextensive systems, the additivity postulate $n^o$4 should
become, for two systems with $N_1$ and $N_2$ elements,
respectively :

\begin{equation}
I(N_1\times N_2)=I(N_1)+I(N_2)+f(I(N_1),I(N_2))       \label{24}
\end{equation}
where the form of the function $f(I(N_1),I(N_2))$ is of central
importance for all the rest of this nonextensive information
theory. Considering the necessity to come back to Hartley formula
in a special case and in order to find the easiest way out, we
naturally think of the {\it $q$-deformed logarithmic function}
$\frac{N^{1-q}-1}{1-q}$ which $\rightarrow lnN$ when $q\rightarrow
1$. So I postulate :
\begin{equation}
I(N)=\frac{N^{1-q}-1}{1-q}.                          \label{25}
\end{equation}
This generalized Hartley formula corresponds to the following
postulates:

$n^o1_{nex}$) $I(1)=0$

$n^o2_{nex}$) $I\{[1+(1-q)]^\frac{1}{1-q}\}=1$

$n^o3_{nex}$) $I(N)<I(N+1)$

$n^o4_{nex}$) $I(N_1\times N_2)=I(N_1)+I(N_2)+(1-q)I(N_1)\times
I(N_2)$ (non-additivity or nonextensivity\footnote{This product
form of the nonextensive term is also a consequence of the
existence of thermodynamic equilibrium in the composite
system\cite{Abe01}}).\\
As for postulate $n^o5$, I put

$n^o5_{nex}$) $I(N)=I_w+\sum_{i=1}^w p_i^qI(Ni)$.\\ Notice that,
$p_i$ in the postulate $n^o5$ is replaced in the postulate
$n^o5_{nex}$ by $p_i^q$ due to the incomplete normalization. From
Eq. (\ref{25}) and the postulate $n^o5_{nex}$, we
straightforwardly obtain $I_w$, the information measure given by
an incomplete probability distribution $\{p_1,p_2,...,p_w\}$
defined by $p_i=N_i/N$ :

\begin{equation}
\frac{N^{1-q}-1}{1-q}=I_w+\sum_{i=1}^{w}p_i^q\frac{N_i^{1-q}-1}{1-q}
                                                            \label{25a}
\end{equation}
which yields

\begin{equation}
I_w\propto -\sum_{i=1}^{w}p_i^q\frac{p_i^{1-q}-1}{1-q}, \label{26}
\end{equation}

\section{Nonextensive generalization of BGS statistics}

Now we postulate for the nonextensive entropy :

\begin{equation}
S_q=-k\sum_{i=1}^{w}p_i^q\frac{p_i^{1-q}-1}{1-q} \label{27}
\end{equation}

or

\begin{equation}
S_q= -k\frac{\sum_{i=1}^{w}p_i-\sum_{i=1}^{w}p_i^q}{1-q}
                                                            \label{28}
\end{equation}
or even more simply
\begin{equation}
S_q= k\frac{1-\sum_{i=1}^{w}p_i}{1-q}
                                                            \label{28a}
\end{equation}
where $q>0$ is required by the incomplete normalization Eq.
(\ref{8}). It can be easily verified that all the properties of
Tsallis entropy (nonnegativity, concavity, pseudo-additivity, etc)
\cite{Tsal88} are preserved by this entropy [Eq. (\ref{27}) to
(\ref{28a})] because it is nothing but the Tsallis one with a new
normalization condition.

For {\it microcanonical ensemble}, we extremize $S_q$ with the
condition in equation ($\ref{8}$) and obtain $p_i^q=1/w$ and
\begin{equation}
S_q=k\frac{w^\frac{q-1}{q}-1}{q-1}        \label{28b}
\end{equation}
which tends to $S_1=klnw$ in the $q\rightarrow1$ limit.

For {\it canonical ensemble}, maximum entropy Eq.($\ref{28a}$)
with Eqs. ($\ref{8}$) and ($\ref{8a}$) (for energy) as
constraints, i.e.
\begin{equation}
\delta\left[\frac{S_q}{k}+\frac{\alpha}{1-q}\sum_{i=1}^{w}p_i^q
-\alpha\beta\sum_{i=1}^{w}p_i^qE_i\right]=0
                                                     \label{29}
\end{equation}
yields
\begin{equation}
p_i=\frac{[1-(1-q)\beta(E_i)]^\frac{1}{1-q}}{Z_q}
                                                        \label{30}
\end{equation}
with

\begin{equation}
Z_q=\left[\sum_{i}^w[1-(1-q)\beta
E_i]^\frac{q}{1-q}\right]^\frac{1}{q}.              \label{32}
\end{equation}

To obtain Legendre transformations, we take Eq. (\ref{27}) and
replace $p_i^{1-q}$ by equation ($\ref{30}$), remembering equation
($\ref{8}$) and ($\ref{8a}$), we obtain
\begin{equation}
S_q=k\frac{Z_q^{q-1}-1}{q-1}+k\beta Z_q^{q-1}U_q \label{34}
\end{equation}
which, with the help of the thermodynamic relation
$\frac{1}{T}=\frac{\partial{S_q}}{\partial{U_q}}$, leads to
\begin{equation}
\beta=\frac{Z_q^{1-q}}{kT}                  \label{35}
\end{equation}
and
\begin{equation}
F_q=U_q-TS_q=-kT\frac{Z_q^{q-1}-1}{q-1}.            \label{36}
\end{equation}
The $U_q-Z_q$ relation is a little complicated. From equation
($\ref{8a}$) and ($\ref{30}$), it can be recast as follows
\begin{equation}
U_q=\frac{1}{Z_q^q}\frac{\partial}{\partial{\beta}}Z'_q
                                                        \label{37}
\end{equation}
where $Z'_q$ is given by
\begin{equation}
Z'_q=\sum_{i}^w[1-(1-q)\beta E_i]^\frac{1}{1-q}. \label{38}
\end{equation}

As in Tsallis' case, it is straightforward to verify that all
above relations reduce to those of BGS case in the $q\rightarrow1$
limit.

Now we will discuss some points concerning the nonextensivity of
the system. The generalized Hartley formula Eq. (\ref{25}) or the
nonadditivity postulate $n^o4_{nex}$ suggests that, for two
subsystems $A$ and $B$ of a system $C=A+B$ :

\begin{equation}
N_{ij}(C)=N_i(A)\times N_j(B).                        \label{39}
\end{equation}
and $N(C)=N(A)\times N(B)$. These relations assume the
factorization of the joint probability $p_{ij}$ or $p_{ij}^q$ :

\begin{equation}
p_{ij}^q(C)=p_i^q(A)p_j^q(B)                     \label{40}
\end{equation}
which in turn leads to the nonextensivity of entropy,

\begin{equation}
S_q(A+B)=S_q(A)+S_q(B)+\frac{q-1}{k}S_q(A)S_q(B).
                                                            \label{41}
\end{equation}

Eq. (\ref{40}) is in fact the definition of the independence of
the effective probability $p_i^q(A)$ or $p_i^q(B)$. Considering
the distribution Eq. (\ref{30}), we easily get
\begin{equation}
E_{ij}(A+B)=E_i(A)+E_j(B)+(q-1)\beta E_i(A)E_j(B) \label{42}
\end{equation}
and
\begin{equation}
U_q(A+B)=U_q(A)+U_q(B)+(q-1)\beta U_q(A)U_q(B).   \label{43}
\end{equation}

Eqs. (\ref{40}), (\ref{42}) and (\ref{43}) tell us that if
$p_i(A)$ and $p_i(B)$ are independent, the two systems $A$ and $B$
are dependent on each other and correlated by Eqs. (\ref{42}) or
(\ref{43}). But for two independent systems $A$ and $B$ with
$E_{ij}(C)=E_i(A)+E_j(B)$, we loss Eq. (\ref{40}) and, strictly
speaking, can no more find the relation between $U_q(C)$, $U_q(A)$
and $U_q(B)$, unless we put $q=1$ and come back to BGS case. As
for this problem of correlation, Abe \cite{Abe99} has studied
N-body problem with ideal gas model. He concluded that, in
thermodynamic limits or in the limit of big $N$ (particle number),
the correlation term in Eq. (\ref{42}) and (\ref{43}) would
vanish. The suppression of the correlation, in addition, allows to
establish the zeroth law of thermodynamics within the framework of
nonextensive statistical mechanics with escort
probability\cite{Abe99}. We mention here that the reader can find
an exact establishment of the zeroth law based on Eq. (\ref{41})
and (\ref{43}) without neglecting the energy
correlation\cite{Wang01}.

\section{Conclusion}

In conclusion, the conventional BGS statistical mechanics is
generalized on the basis of the idea that we sometimes can not
know all the possible physical states or the exact probability
distribution of a complicated physical system and so that we have
to use the suitable information theory for incomplete probability
distribution. The most important step of this generalization is
the {\it incomplete normalization} $\sum_i p_i^q=1$ with a free
parameter $q$ (positive) which is dependent on the neglected
interactions. We have seen that, apart from some minor
differences, the present incomplete scenario of nonextensive
statistics has the same characteristics as Tsallis scenario for
complete probability distribution. On the other hand, we would
like to indicate here that, with the hypothesis of incomplete
distribution, the parameter $q$ is logically related to the
quantity $Q$ in Eq.(\ref{6}) and so to the interactions neglected
in the hamiltonian of the system. For example, for a
microcanonical ensemble, we can write : $q=1-\frac{lnQ}{lnp}$ or
$Q=w^{\frac{q-1}{q}}$. In addition, $q$ can be related to internal
energy or to other basic quantities of physical systems
\cite{Wang00}. So the understanding of nonextensive
thermostatistics under the angle of incomplete information may be
of interest for interpreting particular values of $q$ for
different complicated physical
systems\cite{Krog00,Nasc00,Nott66,Ahme00}.

\section{Acknowledgments}

I acknowledge with great pleasure the very useful discussions
with Professors Constantino Tsallis and Alain Le M\'ehaut\'e on
some points of this work. Thanks are also due to Professor M.S. El
Naschie, Dr. Laurent Nivanen, Dr. Fran\c{c}ois Tsobnang and Dr.
Michel Pezeril for valuable comments.

\newpage

{\Large Figure captions :}

Figure 1) Variation of nonextensive microcanonical entropy $S_q$
with countable state number $w$ for different $q$ value. We see
that $S_q$ increases with increasing $q$.

\end{document}